\documentclass[fleqn,usenatbib]{mnras}

\usepackage{newtxtext,newtxmath}

\usepackage[T1]{fontenc}
\usepackage{ae,aecompl}

\usepackage{graphicx}	
\usepackage{amsmath}	
\usepackage{hyperref}

\def\simeq{
\mathrel{\raise.3ex\hbox{$\sim$}\mkern-14mu\lower0.4ex\hbox{$-$}}
}

\def\ltsima{$\; \buildrel < \over \sim \;$}
\def\simlt{\lower.5ex\hbox{\ltsima}}
\def\gtsima{$\; \buildrel > \over \sim \;$}
\def\simgt{\lower.5ex\hbox{\gtsima}}

\def\msun{{\,\rm M_{\odot}}}

\def\be{\begin{equation}}
\def\ee{\end{equation}}

\def\del#1{{}}
\def\ltsima{$\; \buildrel < \over \sim \;$}
\def\simlt{\lower.5ex\hbox{\ltsima}}
\def\gtsima{$\; \buildrel > \over \sim \;$}
\def\simgt{\lower.5ex\hbox{\gtsima}}
\def\sgra{Sgr~A$^*$}

\def\Pon{P\=oniu\=a'ena }

\title[High-z SMBHs can grow from stellar seeds]{High-redshift SMBHs can grow from stellar-mass seeds via chaotic accretion}

\author[K. Zubovas, A. R. King]{Kastytis Zubovas$^{1,2,\star}$ and Andrew King$^{3,4,5}$ \\
  $^{1}$Center for Physical Sciences and Technology, Saul\.{e}tekio al. 3, Vilnius LT-10257, Lithuania\\
  $^{2}$Astronomical Observatory, Vilnius University, Saul\.{e}tekio al. 3, Vilnius LT-10257, Lithuania\\
  $^{3}$Department of Physics \& Astronomy, University of Leicester, Leicester, LE1 7RH, UK \\
  $^{4}$ Astronomical Institute Anton Pannekoek, University of Amsterdam, Science Park 904, 1098 XH Amsterdam, Netherlands\\ 
  $^{5}$ Leiden Observatory, Leiden University, Niels Bohrweg 2, NL-2333 CA Leiden, Netherlands \\
  $^{\star}$ {E-mail:~} {\rm kastytis.zubovas@ftmc.lt} }

\date{Accepted XXX. Received YYY; in original form ZZZ}

\pubyear{2019}

\begin{document}
\label{firstpage}
\pagerange{\pageref{firstpage}--\pageref{lastpage}}
\maketitle

\begin{abstract}

Extremely massive black holes, with masses $M_{\rm BH} > 10^9 \msun$, have been observed at ever higher redshifts. These results create ever tighter constraints on the formation and growth mechanisms of early black holes. Here we show that even the most extreme black hole known, \Pon\!, can grow from a $10 \msun$ seed black hole via Eddington-limited luminous accretion, provided that accretion proceeds almost continuously, but is composed of a large number of episodes with individually-uncorrelated initial directions. This chaotic accretion scenario ensures that the growing black hole spins slowly, with the dimensionless spin parameter $a \simlt 0.2$, so its radiative efficiency is also low, $\epsilon \simeq 0.06$. If accretion is even partially aligned, with $20-40\%$ of accretion events happening in the same direction, the black hole spin and radiative efficiency are much higher, leading to significantly slower growth. We suggest that the chaotic accretion scenario can be completely falsified only if a $10^9 \msun$ black hole is discovered at $z \geq 9.1$, approximately $150$~Myr before \Pon\!. The space density of extreme quasars suggests that only a very small fraction, roughly one in $4 \times 10^7$, of seed black holes need to encounter favourable growth conditions to produce the observed extreme quasars. Other seed black holes grow much less efficiently, mainly due to lower duty cycles, so are much more difficult to detect.

\end{abstract}




\section{Introduction} \label{sec:intro}

It is now generally accepted that most galaxies harbour supermassive black holes (SMBHs) in their centres \citep{Merloni2013book, Graham2016ASSL}. This is true both for galaxies in the Local Universe and at high redshift, where almost 200 quasars have been discovered at $z>6$ \citep{Willott2010AJ, Carnall2015MNRAS, Matsuoka2016ApJ, Banados2016ApJS, Jiang2016ApJ, Mazzucchelli2017ApJ, Reed2017MNRAS, Wang2017ApJb, Matsuoka2018PASJ, Matsuoka2018ApJS, Reed2019MNRAS, Wang2019ApJ, Vito2019A&A}. These quasars are typically powered by SMBHs with masses of order $10^9 \msun$, yet are seen at lookback times $> 13$~Gyr, i.e. when the Universe was less than $800$~Myr old. The most extreme SMBH currently known is \Pon with a mass of $M_{\rm Pon} = 1.5 \times 10^9 \msun$, seen at $z = 7.515$ \citep{Yang2020ApJ}, corresponding to $t = 700$~Myr after the Big Bang\footnote{All conversions between redshift and time in this paper were made using Ned Wright's Cosmology Calculator \citep{Wright2006PASP}, available at \url{http://www.astro.ucla.edu/~wright/CosmoCalc.html}}. These observations provide a rather tight constraint on the possible models of the origin and early growth of SMBHs.

In general, these models fall broadly into two groups \citep[see][for a review]{Inayoshi2019arXiv}. One group considers SMBHs growing from massive seeds: objects with masses $\sim10^4-10^5 \msun$ \citep{Ferrara2014MNRAS} that form via direct gas collapse \citep{Loeb1994ApJ, Begelman2006MNRAS, Dijkstra2014MNRAS, Chon2016ApJ, Wise2019Natur}, an intermediate supermassive star stage \citep{Regan2009MNRAS, Volonteri2010MNRAS, Begelman2010MNRAS} or due to rapid mergers of stars and/or individual stellar-mass black holes in a dense cluster \citep{PortegiesZwart2002ApJb, PortegiesZwart2004Natur, Devecchi2009ApJ}. These seeds can then grow the required four orders of magnitude in mass over several hundred Myr even if the radiative efficiency of accretion is $\epsilon \simgt 0.1$.

The second kind of model suggests that SMBHs grow from stellar-mass seeds \citep{Madau2001ApJ} via luminous accretion with a rather low radiative efficiency. Low efficiency may be achieved if accretion rates are hyper-Eddington; they create conditions where the photons emitted by the accreting gas cannot escape the accretion flow and are dragged into the black hole \citep{Alexander2014Sci, Sadowski2015MNRAS, Pacucci2015MNRAS, Inayoshi2016MNRAS, Sakurai2016MNRAS, Takeo2020MNRAS}. Even if accretion consists of Eddington-limited thin disc episodes, a radiative efficiency $\epsilon \sim 0.06$ may be achieved if the spin of the black hole is kept low \citep[e.g.][]{King2008MNRAS}. Since the black hole mass depends exponentially on $\left(1-\epsilon\right)/\epsilon$, this difference is enough to allow for the growth of SMBHs powering even the most extreme high-redshift quasars.

The evolution of the SMBH spin during a series of accretion episodes depends on how the disc aligns with the hole. Early calculations \citep[e.g.][]{Thorne1974ApJ,Scheuer1996MNRAS} suggested that the disc always aligns stably, and so the hole should be rapidly spun up to almost the maximum possible spin, $a \simeq 0.998$, where $a$ is the dimensionless spin parameter, with $a = 0$ representing a Schwarzschild black hole and $a = 1$ representing a maximally-spinning Kerr black hole. However, the conclusion of stable alignment rests on the erroneous implicit assumption that the disc angular momentum dominates over that of the hole; in reality, the opposite is often the case, and so stable counter-alignment may occur during many accretion episodes \citep{King2005MNRAS}. In these cases, accretion reduces SMBH spin \citep{King2005MNRAS}. A single counter-aligned accretion episode reduces the spin much more than a single aligned episode increases it, leading to a tendency for the SMBH spin to decrease over time \citep{King2008MNRAS}, provided that accretion is chaotic, i.e. individual accretion episodes have uncorrelated initial directions of angular momentum. Numerical modelling of the evolution of SMBHs under realistic cosmological conditions \citep[e.g.][]{Griffin2019MNRAS} mostly confirms this idea, although they do not produce large enough populations to test for the presence of extreme high-redshift SMBHs.

In this paper, we investigate whether a chaotic accretion scenario can produce extremely massive SMBHs at $t < 800$~Myr after the Big Bang (corresponding to $z \sim 6.7$). We show that as long as accretion events are not strongly correlated ($<20\%$ of all accretion events happen along the same axis) and assuming a duty cycle close to $100\%$, a \Pon\!-like SMBH can be produced. This suggests that the extreme quasars might be powered by accretion on to SMBHs with rather unique growth histories, but such SMBHs are much more likely to be observed at high redshift. We suggest ways to test our model, in particular by searching for lower-mass and/or higher-redshift SMBHs than the current extreme ones. We also predict that these extreme SMBHs have rather low spins and should be significantly obscured.

The paper is structured as follows. We begin by exploring the constraints on SMBH seed mass and radiative efficiency provided by the observations of extreme quasars in Section \ref{sec:constraints}. In Section \ref{sec:model}, we briefly describe the chaotic accretion model and its numerical implementation. In Section \ref{sec:results}, we present the results of numerical calculations of SMBH growth. We discuss their implications in Section \ref{sec:discuss} and summarize in Section \ref{sec:conclusion}.

Throughout the paper, we use the most recent Planck cosmology \citep{Planck2020AA}: Hubble parameter $H_0=67.4\pm0.5$~km~s$^{-1}$~Mpc$^{-1}$, matter density parameter $\Omega_{\rm m} =0.315\pm0.007$ and dark energy density parameter $\Omega_\Lambda \equiv 1 - \Omega_{\rm m}$.

\section{Constraints on SMBH origin and growth} \label{sec:constraints}

\begin{table*}
	\centering
	\caption{Properties of a few selected $z > 6$ SMBHs. Objects are ordered by descending strength of constraint on formation and growth mechanisms (see Figure \ref{fig:constraints}). Where two references exist, the first is the discovery paper and the second provides the mass estimate used.}
	\label{tab:constraints}
	\begin{tabular}{lccc} 
		\hline
		Object name & redshift & log$\left(M_{\rm BH}/\msun\right)$ & Reference\\
		\hline
		\Pon            & $7.515$   & $9.18$ & \cite{Yang2020ApJ} \\
		ULASJ1342+0928  & $7.5413$  & $8.89$ & \cite{Banados2018Natur} \\
		ULASJ1120+0641  & $7.0842$  & $9.39$ & \cite{Mortlock2011Natur}, \cite{Mazzucchelli2017ApJ} \\
		SDSSJ0100+2802  & $6.3258$  & $10.03$& \cite{Wu2015Natur} \\
		PSOJ338+29      & $6.666$   & $9.43$ & \cite{Venemans2015ApJ}, \cite{Mazzucchelli2017ApJ} \\
		SDSSJ1148+5251  & $6.4189$  & $9.71$ & \cite{Fan2003AJ}, \cite{DeRosa2011ApJ} \\
		VIKJ0109-3047   & $6.7909$  & $9.12$ & \cite{Venemans2013ApJ}, \cite{Mazzucchelli2017ApJ} \\
		SDSSJ2310+1855  & $6.0031$  & $9.62$ & \cite{Wang2013ApJ}, \cite{Jiang2016ApJ} \\
		\hline
	\end{tabular}
\end{table*}

\begin{figure}
	\includegraphics[width=\columnwidth]{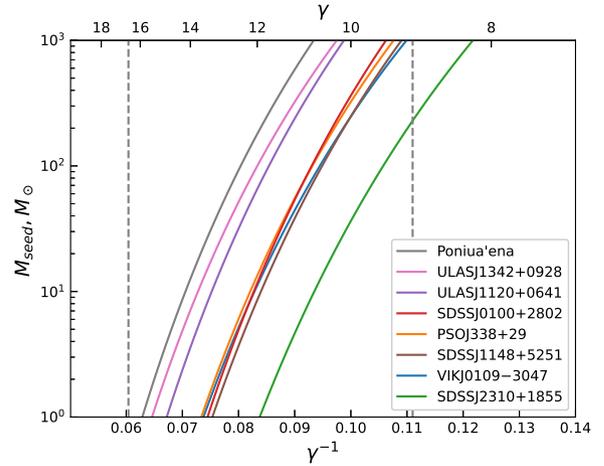}
    \caption{The relationship between seed black hole mass $M_{\rm seed}$ and growth efficiency $\gamma$ for a few selected $z > 6$ SMBHs (see Table \ref{tab:constraints} for data and references). We use $\gamma^{-1}$ as the independent variable for easier comparison with the closely related radiative efficiency $\epsilon$.}
    \label{fig:constraints}
\end{figure}

Supermassive black holes gain most of their mass via luminous accretion \citep{Soltan1982MNRAS, Yu2002MNRAS, Marconi2003ApJ, Shankar2004MNRAS, Hopkins2006ApJ}. The mass of an SMBH cannot grow much faster than the Eddington-limited growth rate
\begin{equation} \label{eq:mt}
    M_{\rm BH}\left(t\right) = M_{\rm seed} \times {\rm exp}\left(\frac{t-t_{\rm form}}{t_{\rm Sal}}\delta l_{\rm E}\right).
\end{equation}
Here, $M_{\rm seed}$ is the mass of the `seed' black hole formed at $t = t_{\rm form}$, and $t_{\rm Sal}$ is the Salpeter timescale, given by
\begin{equation}
    t_{\rm Sal} = \frac{\kappa_{\rm e.s.} c}{4 \pi G} \frac{\epsilon}{1-\epsilon} \simeq 45 \frac{\epsilon}{0.1} \frac{0.9}{1-\epsilon} {\rm Myr},
\end{equation}
where $\kappa_{\rm e.s.} \simeq 0.346$~cm$^2$~g$^{-1}$ is the electron scattering opacity, $\epsilon$ is the radiative efficiency of accretion, $c$ is the speed of light and $G$ is the gravitational constant. The factor $\delta$ is the duty cycle of luminous accretion and $l_{\rm E} \equiv \langle L_{\rm AGN}/L_{\rm Edd}\rangle$ is the average Eddington factor.

Equation (\ref{eq:mt}) has five free parameters: $M_{\rm seed}$, $t_{\rm form}$, $\epsilon$, $\delta$ and $l_{\rm E}$. However, the last three can be combined into a single ``growth efficiency'' $\gamma =  \delta l_{\rm E} \left(1-\epsilon\right)\epsilon^{-1}$. The seed formation time can be estimated, at least for seeds forming from gas clouds, via numerical simulations and observations. Numerical simulations show that star formation begins at $z \sim 30$, corresponding to $t = 100$~Myr after the Big Bang \citep{Johnson2008MNRAS}. Observationally, the imprint of the UV radiation from the first stars on the cosmic microwave background becomes evident at $z \sim 20$, corresponding to $t = 180$~Myr \citep{Bowman2018Natur}. Using this latter estimate of the formation timescale is rather conservative, since occasional star formation may have happened significantly earlier, but it would have been too scattered, or too obscured, to produce a detectable signal. Due to this uncertainty, we take $t_{\rm form} = 100$~Myr in the rest of this paper, unless noted otherwise.

For any given SMBH observed at any particular time, there is a unique relationship between $M_{\rm seed}$ and $\gamma$ that can be derived from equation (\ref{eq:mt}). We show these relationships for eight extreme $z > 6$ SMBHs in Figure \ref{fig:constraints}\footnote{All plots are made using the Matplotlib Python package \citep{matplotlib}.}. The chosen SMBHs include the most distant \citep[ULASJ1342+0928,][]{Banados2018Natur}, most massive \citep[SDSSJ0100+2802,][]{Wu2015Natur} and the most constraining (\Pon\!) ones; table \ref{tab:constraints} gives the list of objects, their redshifts and masses. The vertical dashed line on the left of the diagram highlights $\gamma \simeq 0.0604^{-1} \simeq 16.5$, which is the highest possible growth efficiency: this value represents an extreme combination $\epsilon = 0.057$, appropriate for a non-spinning SMBH, together with $\delta = l_{\rm E} = 1$. The other vertical dashed line highlights $\gamma \simeq 0.111^{-1} \simeq 9$, which for a $\delta = l_{\rm E} = 1$ case corresponds to $\epsilon = 0.1$, typically assumed in many models of SMBH growth.

It is immediately clear that the value of growth efficiency has an enormous effect on the constraint of $M_{\rm seed}$. An increase from $\gamma = 9$ to $\gamma = 11.5$, corresponding (for the extreme feeding case) to $\epsilon$ decrease from $0.1$ to $0.08$, reduces the required seed mass by a factor $\sim 40$, bringing even the most extreme quasars within reach of growth models based on almost continuous accretion on to stellar mass seeds. This change in growth efficiency falls well within the range of uncertainty in observational constraints of $\epsilon$ \citep[e.g.][]{Shankar2004MNRAS, Shankar2009ApJ, Davis2011ApJ, Pacucci2015MNRAS, Zhang2017SCPMA, Davies2019ApJ, Shankar2020NatAs} and theoretical studies \citep{Tucci2017AA}.

\section{Chaotic accretion} \label{sec:model}

The model we use follows the one presented in \cite{King2005MNRAS, King2006MNRAS, King2007MNRAS, King2008MNRAS}. Here we only present the basic qualitative arguments and refer the reader to those papers for more details.

A SMBH grows via a series of individual accretion episodes. The maximum mass that can be accreted during a single episode is given by the self-gravity condition \citep[cf.][]{Pringle1981ARAA}
\begin{equation} \label{eq:deltam1}
    \Delta M_{\rm BH,1} = \left(1-\epsilon\right)M_{\rm disc} \leq \left(1-\epsilon\right)\frac{H}{R}M_{\rm BH},
\end{equation}
where $H/R$ is the disc aspect ratio and the factor $1-\epsilon$ represents the fraction of disc mass lost to radiation. The radiative efficiency depends monotonously on the dimensionless spin parameter $a$: for a disc counter-aligned with a maximally spinning black hole, $a = -1$ and $\epsilon \simeq 0.038$; for a non-spinning (Schwarzschild) black hole, $a = 0$ and $\epsilon \simeq 0.057$; and for prograde alignment between the disc and a maximally spinning black hole, $a = 1$ and $\epsilon \simeq 0.42$ \citep[cf.][]{Bardeen1970Natur, King2006MNRAS}.

Each accretion episode initially forms a disc with an angular momentum $J_{\rm d}$ at some angle $\theta$ to the direction of the hole's angular momentum $J_{\rm h}$. \cite{King2005MNRAS} showed that the disc can align stably with the black hole in either a prograde or retrograde fashion, with the latter case occuring if $\theta$ satisfies a condition
\begin{equation} \label{eq:costheta}
    {\rm cos}\theta < - \frac{J_{\rm d}}{2J_{\rm h}}.
\end{equation}
Since the disc mass is always much smaller than the hole mass, as long as the hole is not spinning very slowly, stable counter-alignment might occur in close to half of all accretion episodes.

As the accretion disc evolves, most of its angular momentum is transferred to large radii and eventually escapes the system \citep{Shakura1973AA, Pringle1981ARAA}. The angular momentum actually transferred to the SMBH is
\begin{equation} \label{eq:jacc}
    J_{\rm acc} = \pm \Delta M_{\rm BH,1} j_{\rm ISCO}.
\end{equation}
The specific angular momentum, $j_{\rm ISCO}$ is the angular momentum of gas per unit mass at the innermost stable circular orbit. The sign of $J_{\rm acc}$ is positive if the disc is aligned with the SMBH and negative if it is counter-aligned.

Over a large number of individual accretion episodes, the SMBH spin evolves in a stochastic fashion. If the episodes are uncorrelated, i.e. the initial direction of $\vec{J}_{\rm d}$ is random for each episode, the trend is for the spin to decrease, because a counter-rotating disc has a significantly greater ISCO radius ($R_{\rm ISCO} = 9 GM_{\rm BH}/c^2$ for a maximally spinning black hole) than a prograde disc ($R_{\rm ISCO} = GM_{\rm BH}/c^2$ for a maximally spinning black hole); as a result, the accreted angular momentum is much greater in counter-aligned accretion events. On the other hand, if the accretion episodes are correlated to some extent, the SMBH gradually begins to spin in the same direction and its spin value increases.

\subsection{Numerical model}

\begin{figure}
	\includegraphics[width=\columnwidth]{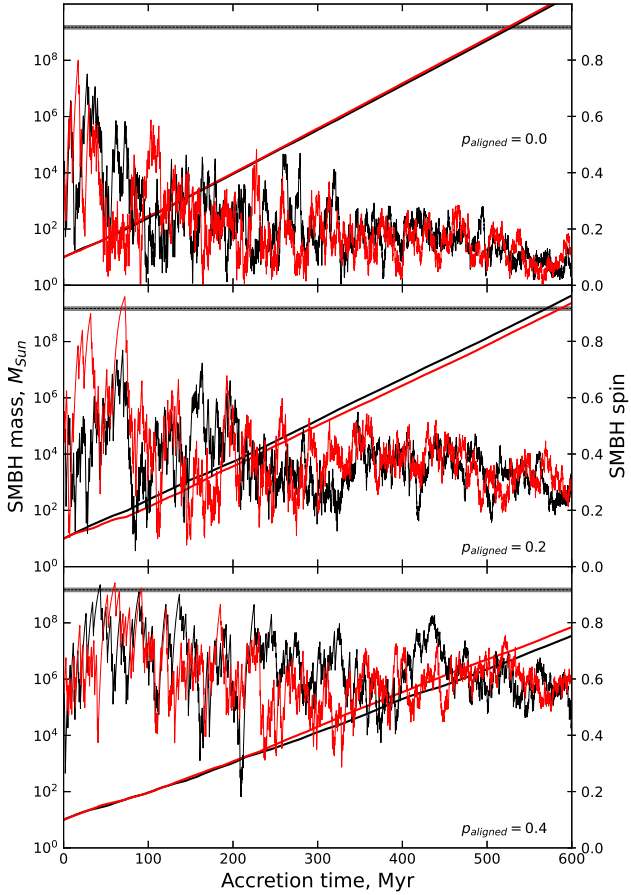}
    \caption{Evolution of SMBH mass (almost straight lines, scale on the left) and spin (jagged lines, scale on the right) for two example growth scenarios in simulations with completely chaotic accretion (top) and simulations with $20\%$ (middle) or $40\%$ (bottom) of accretion events are aligned in the z direction. Horizontal dashed lines show the mass of P\=oniu\=a'ena and the shaded region represents its uncertainty.}
    \label{fig:growth_examples}
\end{figure}

\begin{figure}
	\includegraphics[width=\columnwidth]{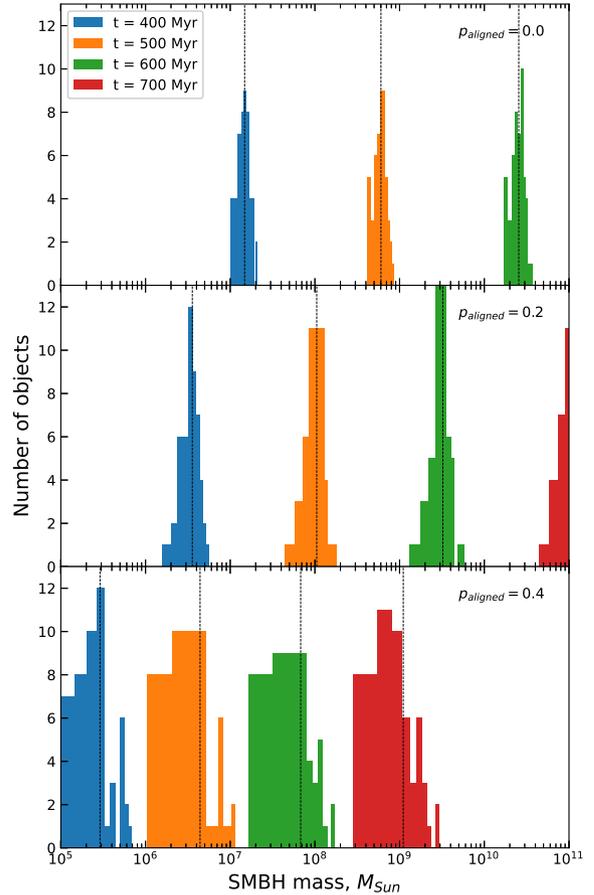}
    \caption{Histograms of SMBH masses at $t = 400, 500, 600$ and $700$~Myr of accretion (blue, orange, green and red, respectively) in simulations with completely chaotic accretion (top) and simulations with $20\%$ (middle) or $40\%$ (bottom) of accretion events are aligned in the z direction. Each histogram comprises 50 mass values obtained by a different set of randomised accretion events. Vertical dashed lines show the mean of each histogram. The rightmost histogram of completely chaotic accretion is not visible because all masses are $> 10^{11} \msun$.}
    \label{fig:mass_histograms}
\end{figure}

We test the evolution of black hole mass and spin, as well as the final SMBH mass distributions, under various accretion alignment conditions, using a numerical model. The model is initialised with a seed black hole of mass $M_0 = 10 \msun$, with a random spin magnitude and direction. A series of accretion events are then used to grow the SMBH. Each event is limited in mass by the self-gravity condition (eq. \ref{eq:deltam1}); we also checked that imposing an additional maximum mass of an individual event, to simulate the possible mass limit of gas clouds feeding the SMBH, has no effect on our results. The initial direction of the disc's spin is chosen completely randomly in one model; in two others, we add an additional parameter $p_{\rm aligned} = 0.2; 0.4$, and randomly select this fraction of accretion events to have spins aligned in the positive-z direction. The code then checks the condition for hole-disc alignment (eq. \ref{eq:costheta}) and calculates the new direction of the hole's angular momentum by considering the vector sum of $\vec{J_{\rm d}}$ and $\vec{J_{\rm h}}$. The mass and angular momentum of the black hole are updated by adding the appropriate fractions of the disc mass (eq. \ref{eq:deltam1}) and angular momentum (eq. \ref{eq:jacc}). The new angular momentum is used to calculate the spin parameter and the radiative efficiency to be used for the next accretion event. The duration of a single episode is calculated as the time for the AGN to consume the disc mass at the given luminosity:
\begin{equation}
    t_{\rm ep} = \frac{\epsilon M_{\rm disc} c^2}{L_{\rm AGN}}.
\end{equation}
We set the luminosity to be equal to the Eddington luminosity of the black hole. While the luminosity is likely to vary a lot during SMBH growth, we are mainly interested in the early Universe and in extreme black holes which are likely to have grown in dense gas environments where prolonged feeding at the Eddington (or higher) rate was common. If the black hole is fed at a mildly super-Eddington rate, the accretion disc sheds material at radii larger than ISCO and the black hole does not grow faster than at the Eddington rate \citep{Shakura1973AA, Pringle1981ARAA}.

The code runs, updating the SMBH mass and spin and advancing the time, until either $700$~Myr have passed or the SMBH mass exceeds $10^{11} \msun$. We implicitly assume that the SMBH mass and spin do not change in between accretion events, so the time tracked in the model is actually the ``accretion time'', defined over sufficiently long timescales as
\begin{equation}
    t_{\rm acc} = \delta t,
\end{equation}
where $t$ is actual time elapsed. We run 150 simulations in total, 50 each with $p_{\rm aligned} = 0$, $0.2$ and $0.4$. The only differences between the simulations in each set are the random numbers used to define the initial SMBH spin magnitude and direction and the directions of individual accretion episodes.

\section{Black hole growth: results} \label{sec:results}

In Figure \ref{fig:growth_examples}, we present the mass (straight lines, logarithmic scale on the left) and spin (jagged lines, scale on the right) evolution for two randomly-selected black holes in purely chaotic accretion scenarios (top panel) and scenarios with $p_{\rm aligned} = 0.2$ (middle) and $p_{\rm aligned} = 0.4$ (bottom). The horizontal line indicates the mass on \Pon at $M_{\rm Pon} = 1.5 \times 10^9 \msun$ and the shaded region corresponds to the uncertainty in this estimate.

The purely chaotic accretion scenario evolves in the same way as was shown previously by \cite{King2008MNRAS}. The black hole spin gradually decreases and oscillates in the range $0 < a < 0.2$, while the mass grows exponentially and reaches the mass of \Pon in $t \sim 520$~Myr. The different directions of individual accretion events manifest in the evolution of spin, while the mass growth is barely affected. Models with partially aligned accretion events show higher SMBH spins and, as a result, slower mass growth. In the model with $p_{\rm aligned} = 0.2$, spin oscillates in the range $0.2 < a < 0.4$ and the mass reaches that of \Pon after $t = 570-590$~Myr of accretion. Finally, in the model with $p_{\rm aligned} = 0.4$, spin ends up in the range $0.5 < a < 0.7$ and it takes $\sim 700-730$~Myr to grow the SMBH to $1.5\times 10^9 \msun$.

Despite the difference in individual accretion episode directions, both the overall spin and mass evolutions show remarkably similar trends. However, with more alignment, the difference in SMBH mass between the two example models at a given time increases. In order to show these differences in more detail, as well as the range of SMBH masses we may expect at different times in different scenarios, we plot, in Figure \ref{fig:mass_histograms}, the histograms of SMBH masses in the three models at $t_{\rm acc} = 400$, $500$, $600$ and $700$~Myr accretion time. Each histogram is produced by running the model 50 times with different sets of random numbers determining the initial disc angular momentum direction. The vertical dotted lines show the mean value of each histogram.

The variance of final SMBH masses grows with increasing alignment of accretion events. In the purely chaotic case (top panels), the difference between the most- and least-massive SMBHs is a factor $2$, while in the most aligned case (bottom panels), it is a factor $10$. This happens because when the SMBH spin is higher, radiative efficiency is more sensitive to spin, leading to small changes in spin value resulting in more significant changes in SMBH growth rate. These variations, are, of course, only lower limits on the actual spread of SMBH masses at any given time, because we do not account for the different duty cycles of individual black holes.

\section{Discussion}  \label{sec:discuss}

Our results show that it is possible to grow a black hole from a stellar-mass seed to the size of \Pon in $t \sim 520$~Myr, i.e. rapidly enough to produce the black hole assuming that the first stars formed $180$~Myr after the Big Bang \citep{Bowman2018Natur} or earlier. Such a scenario rests on several implicit assumptions and produces several predictions regarding the properties of high-redshift extreme SMBHs and their effect on the environment. We consider them in turn.

\subsection{Masses and alignment of individual accretion discs} \label{sec:discsize}

The time coordinate in our models is the accretion time rather than total time of the black hole's existence. So the fact that $t_{\rm acc} \sim 520$~Myr is required for production of a \Pon-mass SMBH means that the duty cycle of its growth has to be almost $100\%$. This may be unrealistic combined with the necessity of having a low rate of alignment of accretion directions; in fact, one might expect accretion directions to be highly correlated if accretion is essentially continuous. Nevertheless, there are several ways of mitigating this tension.

The masses of individual accretion episodes in our model are limited by the self-gravity condition \citep{Pringle1981ARAA, King2007MNRAS}. The validity of this condition rests on two assumptions: that the disc is geometrically thin and optically thick, i.e. that the standard thin-disc equations are applicable to describing its structure, and that cooling is efficient at radii $R \simgt 10^3 r_{\rm s}$, where $r_{\rm s} = 2 G M_{\rm BH} / c^2$ is the Schwarzschild radius of the SMBH. Thin discs are the standard solution describing an accretion flow with mass flow rate $0.01 \dot{M}_{\rm Edd} \simlt \dot{M} \simlt 2 \dot{M}_{\rm Edd}$ \citep{Best2012MNRAS, Sadowski2013MNRAS, Inayoshi2016MNRAS}. The importance of self-gravity in the outer regions of these discs has been recognised by numerous authors \citep[e.g.,][]{Shlosman1990Natur, Gammie01, Rice05}. In addition, cooling is typically efficient in these regions and leads to star formation \citep{Shlosman1989ApJ, Collin1999A&A, Goodman2003MNRAS}; even if the forming stars heat the disc up and prevent further fragmentation, this does not shut off the growth of these stars \citep{Nayakshin2006MNRASb}. The stellar rings around \sgra\!, the central SMBH of the Milky Way, provide strong evidence for the validity of this scenario \citep{Paumard2006ApJ, Bonnell2008Sci, Bartko2009ApJ}. Cooling in primordial accretion discs is probably less efficient due to low metallicity, but this is unlikely to stabilize them significantly against self-gravity \citep{Tanaka2014MNRAS}.

\subsection{Gas reservoir mass and accretion duty cycle} \label{sec:reservoir}

The gas reservoir surrounding the SMBH is likely to be turbulent, perhaps composed of streams and orbiting clumps. In this case, the streams of gas approaching the SMBH may easily have different directions, even if the reservoir as a whole has some angular momentum. The radius of the feeding reservoir immediately available to the SMBH is $\sim R_{\rm infl}$, the radius of influence of the black hole's gravity \citep{Dehnen2013ApJ}, which is much higher than the maximum radius of a non-self-gravitating accretion disc. As a result, even a small difference in the impact parameter of two streams falling toward the SMBH will lead to significant differences in the angle of the forming disc \citep{Nayakshin2012ApJ}. This suggests that even essentially continuous accretion can be almost entirely chaotic.

Streams falling in from various directions also continuously change the orientation of the accretion disc. This happens on approximately the dynamical timescale of the reservoir,
\begin{equation}
    t_{\rm d,res} \sim \frac{R_{\rm res}}{\sigma} \sim 5\times10^5 R_{100} \sigma_{200}^{-1} {\rm yr},
\end{equation}
where we scale the reservoir size to 100~pc and the velocity dispersion to 200~km~s$^{-1}$. This should have little effect on the initial alignment between the disc and the hole, because only the warped region of the disc is affected by this. So long as $M_{\rm BH} \simlt 10^8 \msun$, the warp radius is smaller than the self-gravity radius and the initial alignment between the black hole and the disc is not affected by the outer regions of the disc.

Some galaxies have disky structures surrounding their central SMBHs on scales of order $0.1-1$~pc, e.g. NGC 4258 \citep{Herrnstein2005ApJ}, IC 2560 \citep{Yamauchi2012PASJ} and NGC 1068 \citep{Imanishi2020ApJ}. Although these are almost certainly strongly self-gravitating systems \citep[e.g.,][]{Kumar1999ApJ, Lodato2003A&A, Martin2008MNRAS, Tilak2008ApJ}, we can consider, for the sake of argument, what happens if such systems are prevented from fragmenting. Over time, the accretion disc aligns with the black hole in a prograde fashion. However, this this happens on the viscous timescale at the radius where the disc angular momentum starts to dominate over that of the hole, which is \citep[cf.][]{King2005MNRAS}
\begin{equation}
    t_{\rm al} \sim 2\times 10^7 M_8^{6/19} \left(\frac{a}{0.2}\right)^{14/19} {\rm yr},
\end{equation}
where $M_8$ is the black hole mass scaled to $10^8 \, \msun$ and assuming Eddington-rate accretion. The ratio between the two timescales is
\begin{equation}
    \frac{t_{\rm d,res}}{t_{\rm al}} \sim 0.025 R_{100} \sigma_{200}^{-1} M_8^{-6/19} \left(\frac{a}{0.2}\right)^{-14/19}.
\end{equation}
This ratio is much lower than unity except for the smallest black holes, and even then only if the feeding reservoir is as large as 100~pc. If the black hole is initially fed inside a cloud with radius $R_{\rm cl} \sim 24$~pc, the dynamical timescale is shorter than the alignment timescale even for a $10 \, \msun$ seed black hole. As a result, new gas streams effectively randomize the direction of the accretion disc much faster than the disc can cause the SMBH to align with it.

The reservoir is depleted by accretion and small-scale momentum-driven feedback, but outflows are unlikely to affect dense gas outside of the sphere of influence for as long as the SMBH mass is lower than the critical $M_\sigma$ mass \citep{King2010MNRASa}. The depleted reservoir is continuously refilled by gas falling in from larger distances throughout the host halo. Assuming that the halo is gas-rich, as appropriate for high redshift, the infall rate may be a significant fraction of the dynamical rate \citep[cf.][]{Dehnen2013ApJ}
\begin{equation}
    \dot{M}_{\rm dyn} = f_{\rm g} \frac{\sigma^3}{G} \simeq 300 \frac{f_{\rm g}}{0.16} \sigma_{200}^3 \msun {\rm yr}^{-1},
\end{equation}
where we scale the gas fraction $f_{\rm g}$ to the cosmological value $0.16$ and the velocity dispersion $\sigma$ to $200$~km~s$^{-1}$. This rate is much higher than the Eddington accretion rate even for a SMBH with the mass equivalent to \Pon\!. Considering that the most massive black holes should reside in haloes with the highest velocity dispersion, refilling the reservoir surrounding the black hole is even easier.

\subsection{SMBH seeds} \label{sec:seeds}

The seed mass we chose, $M_0 = 10 \msun$, is also rather conservative. It is comparable to that expected for supernova remnants of present-day stars. Early, Population III, stars are expected to be more massive on average than present-day ones \citep{Hirano2014ApJ, Susa2014ApJ, Stacy2016MNRAS}; as a result, their explosion remnants might be more massive as well. The final mass of the black hole at any given time is approximately proportional to the seed mass, so a more massive seed makes it slightly easier to grow an extremely massive SMBH quickly. Furthermore, the effective seed mass might be higher if the black hole forms in a cluster and merges with a few other similar black holes while its mass is still small. However, the constraint on SMBH spin, and hence on the maximum level of alignment, hardly changes if we assume higher seed masses - this can be seen in Figure \ref{fig:constraints}.

Our model is insensitive to the precise redshift of the formation of the black hole seed. In principle, even a seed forming as late as at $z \sim 18$, corresponding to $t = 180$~Myr after the Big Bang, can give rise to \Pon\!, provided that it maintains a duty cycle $\delta = 1$, so that $t_{\rm acc} = t$. If the seed forms at $t = 100$~Myr, a duty cycle of $520/600 \simeq 0.87$ is enough.

\subsection{The most distant SMBH possible}

The possibilities described in sections \ref{sec:discsize} - \ref{sec:seeds} allow for some variation of model parameters, but the observational constraints are still tight. They also provide a stringent test for our model. In particular, the model would be falsified if an SMBH would be detected at a time $t_{\rm obs}$ with a mass greater than
\begin{equation}
    M_{\rm max} \sim 100 \, {\rm exp}\left(\frac{t_{\rm obs}-100 \,{\rm Myr}}{27.6 \,{\rm Myr}}\right) \msun.
\end{equation}
This corresponds to $M = 10^9 \msun$ at $t = 545$~Myr after the Big Bang, or, equivalently, $z \simeq 9.1$. The growth timescale of $27.6$~Myr corresponds to the purely chaotic SMBH growth model, and we use the large seed mass $M_0 = 100 \msun$ for this estimate. Quasars powered by such SMBHs - if they exist - should be discovered by upcoming {\em Euclid} \citep{Roche2012MNRAS} and {\em Athena} \citep{Nandra2013arXiv, Aird2013arXiv} missions. Discoveries of slightly less extreme SMBHs, including that of \Pon\!, provide constraints on parameter combinations encompassing the seed black hole mass, duty cycle, formation time and alignment fraction of accretion episodes.

\subsection{SMBH space density} \label{sec:density}

Current observational estimates of the space density of extreme high-redshift quasars with $L > 10^{46}$~erg~s$^{-1}$ are $n_{\rm QSO} \simlt 10^{-7}$~Mpc$^{-3}$ \citep{Kulkarni2019MNRAS, Wang2019ApJ, Shen2020MNRAS}. This luminosity corresponds to the Eddington limit for a $7.7\times10^7\msun$ SMBH. Quasars with bolometric luminosity $L \sim 2\times10^{47}$~erg~s$^{-1}$, corresponding to the luminosity of \Pon \citep{Yang2020ApJ}, are even rarer: $n_{\rm Pon} \simlt 10^{-8}$~Mpc$^{-3}$. Given that we predict an almost 100\% duty cycle for these objects, the space density of SMBHs should be not much higher than that of quasars. These values allow us to estimate the rarity of conditions that give rise to such extreme black holes so quickly. 

The rate of formation of the first stars is estimated at $\dot{M}_* \simeq 2-3\times10^{-5} \msun$~yr$^{-1}$~Mpc$^{-3}$ at $z = 20$ \citep{Johnson2008MNRAS}; this drops down to $\dot{M}_* \simeq 10^{-6} \msun$~yr$^{-1}$~Mpc$^{-3}$ at $z = 30$. These Population III stars were probably more massive, on average, than stars forming today, with an average mass of $\sim 10-100 \msun$ \citep{Hirano2014ApJ, Susa2014ApJ, Stacy2016MNRAS}. Approximately half of the stars may end up forming black holes at the end of their lives \citep{Heger2002ApJ, Hirano2014ApJ}. This leads to the formation rate of seed black holes of $\dot{N}_{\rm seed} \simeq 5\times 10^{-9} - 5\times 10^{-8}$~yr$^{-1}$~Mpc$^{-3}$ at $z = 30$ and $\dot{N}_{\rm seed} \simeq 10^{-7} - 1.5\times 10^{-6}$~yr$^{-1}$~Mpc$^{-3}$ at $z = 20$. Assuming that the seeds of the most extreme quasars must have been produced before $z = 18$ (corresponding to $t = 180$~Myr, leaving $520$~Myr for \Pon to grow), this gives $t_{\rm form} \sim 80$~Myr for seed formation. Over this period, even the most conservative estimate provides a seed BH density $n_{\rm seed} \sim 0.4$~Mpc$^{-3}$, and it is possible that this density was at least an order of magnitude higher. These values are consistent with estimates based on numerical modelling of Population III star formation \citep{Madau2001ApJ, Volonteri2010A&ARv}. Comparing this value with the quasar space density, we see that if only one in $4\times 10^7$ of initial seeds encounters favourable growth conditions, the population of extreme quasars observed at $z \sim 6-7$ is reproduced. 

This result suggests an explanation for the issue of `stellar black hole starvation', which is expected at high redshifts \citep[e.g.][]{Pacucci2015MNRAS, Pacucci2017ApJ, Smith2018MNRAS}. It happens due to two reasons: many nascent stellar black holes can be ejected from their parent star-forming halos \citep{Whalen2012ApJ}, and those that remain cannot accrete material efficiently until the fossil HII region that surrounded the progenitor star cools down \citep{Johnson2007ApJ}. However, a small fraction of black holes can avoid these issues if they are born within a larger protogalactic halo with escape velocity $v_{\rm esc} > 100$~km~s$^{-1}$. A natal kick with $v < v_{\rm esc}$ allows the black hole to escape the fossil HII region very quickly. Small-scale feedback can enhance gas mixing \citep{Dehnen2013ApJ} and suppress fragmentation \citep{Alvarez2009ApJ} around these black holes, promoting accretion at near-Eddington rates. The expected rarity of these objects can also explain why they are not reproduced in numerical and semi-analytical simulations \citep{Smith2018MNRAS, Griffin2019MNRAS}.

We expect that the key parameters determining the growth rate of the SMBH vary smoothly over their allowed range; furthermore, there are no discontinuities or sharp changes in the dependence of the final SMBH mass on these parameters. Due to this, we predict that the mass function of black holes, and of quasars, should be approximately continuous in mass/luminosity, consistent with available observational constraints. However, the slope of the mass function may be very shallow, i.e. the density of less massive black holes might not be much higher than that of the most massive ones. The total X-ray radiation produced by high-redshift quasars suggests that luminous accretion contributed only $< 1000 \msun$~Mpc$^{-3}$ to the SMBH mass density by redshift $z = 6$ \citep{Treister2013ApJ}. The observed quasars already account for $\rho \simeq \langle M_{\rm BH}\rangle n_{\rm QSO} \simeq 100 \msun$~Mpc$^{-3}$, i.e. $>10\%$ of this limit. This means that the smaller black holes cannot have grown very much at these early times. In addition, the lower duty cycle of these smaller black holes means that the {\em quasar} density is significantly lower than the SMBH density. These two factors might explain why the smaller SMBHs at high redshift have not been detected yet.

\subsection{SMBH properties}

Our results provide two important and potentially testable predictions regarding the properties of extreme high-redshift SMBHs: their spins must be low and they should reside in dense gas environments.

Typical spins of black holes growing through purely chaotic accretion settle to $a<0.3$ in $t_{\rm acc} < 350$~Myr. At this time, the black hole mass is only $M \sim 10^6 \msun$. Similarly, if accretion is slightly aligned with $p_{\rm aligned} = 0.2$, black hole spin is $a < 0.5$ when its mass is $10^6 \msun$ or higher. If accretion is aligned more, the black hole does not reach a mass of $10^8 \msun$ over $600$~Myr of accretion. So the most massive black holes, those with $M > 10^8 \msun$, observed when the Universe was $\sim 700$~Myr old or younger, should all have spins lower than $a = 0.5$. Among more moderate black holes, with $10^6 \msun < M < 10^8 \msun$, the spin distribution should be wide, with both low and high spin values possible. Black holes with masses $M < 10^6 
msun$ should generally spin rapidly, although low spin values should be possible in this population as well. This contrasts with available observational constraints on SMBHs in the Local Universe, where black holes with masses $M \simlt 3\times 10^7 \msun$ seem to have spin values close to maximal, and even more massive black holes generally have spins $a > 0.4$ \citep{Vasudevan2016MNRAS, Reynolds2019NatAs, Reynolds2020arXiv}.

There are a few ways of reconciling our prediction with the observationally-estimated spins. First of all, the observed high-redshift quasars represent an extreme population (see Section \ref{sec:density}) and their spins need not be representative of the high-redshift population as a whole. We predict that as SMBHs with more modest masses are discovered at high redshift, and as their spin estimates are obtained, the overall population will come to resemble the locally-observed one more closely, even if not exactly. Secondly, the global mass growth of supermassive black holes is expected to be dominated by mergers since $z \sim 0.5$, i.e. for the past 3-4 Gyr \citep{Bassini2019A&A}. Mergers tend to flatten the spin distribution \citep{Berti2008ApJ}, so even if the extreme quasars have spins representative of the early SMBH population as a whole, the local population will have much more broadly distributed spin values. There is tentative observational evidence, based on the evolution AGN jet power, that the typical spins have been increasing at least since $z = 1$ \citep{Martinez2011MNRAS}. Finally, observational biases lead to easier identification of AGN powered by rapidly spinning SMBHs, and the currently observed population is consistent with a non-negligible number of slowly-spinning black holes \citep{Vasudevan2016MNRAS}. The situation should become significantly clearer once the so-far elusive low mass SMBH population at high redshifts will be discovered, with upcoming observatories such as {\em Lynx} \citep{Gaskin2019JATIS} and {\em Roman} \citep{Green2012arXiv}. In addition, future missions, including {\em Athena} \citep{Barret2019AA} and {\em AXIS} \citep{Mushotzky2019BAAS}, will be able to measure SMBH spins with unprecedented precision even at high redshift, providing better constraints on the evolution of this quantity.

As spin determines radiative efficiency, information about quasar luminosity might also be used to put constraints on black hole spin. The average radiative efficiency over the whole growth period is $\langle \epsilon \rangle \simeq 0.058$ in the purely chaotic models and increases to $\langle \epsilon \rangle \simeq 0.095$ in the $p_{\rm aligned} = 0.4$ models. In order to power the observed luminosity of \Pon\!, $L_{\rm Pon} \simeq 1.9 \times10^{47}$~erg~s$^{-1}$, the SMBH should be accreting material at a rate $\dot{M}_{\rm Pon} \simeq 3.34/\epsilon\, \msun$~yr$^{-1}$. This translates to $57.64\, \msun$~yr$^{-1}$ and $35.2\, \msun$~yr$^{-1}$ for the two alignment cases, respectively. Assuming this represents the rate of spherical infall of fully ionised material at the outer edge of the accretion disc, $r_{\rm out} \simeq 0.01$~pc, the expected column density is
\begin{equation}
\begin{split}
    N_{\rm H} &\simeq \frac{\dot{M}_{\rm Pon}}{4\pi \mu m_{\rm p} \left(2GM_{\rm Pon} r_{\rm out}\right)^{1/2}} \\&\simeq 2.7\times10^{24} \frac{\dot{M}_{\rm Pon}}{50\, \msun \,{\rm yr}^{-1}} \left(\frac{r_{\rm out}}{0.01\,{\rm pc}}\right)^{-1/2} {\rm cm}^{-2},
\end{split}
\end{equation}
where $\mu \simeq 0.5$ is the mean atomic weight per particle and $m_{\rm p}$ is the proton mass. The two cases of low- and high-spin SMBH give expected hydrogen column density around the Compton-thick limit, $N_{\rm H} = 1.5\times10^{24}$~cm$^{-2}$. Future observations might be able to measure the X-ray spectrum of these AGN and determine their Compton thickness, providing constraints on matter infall rates and, hence, the radiative efficiency of accretion.

Our models are consistent with the general expectation of numerical simulations that the most massive SMBHs should be found in dense environments \citep{Overzier2009MNRAS, Costa2014MNRASb, Habouzit2019MNRAS}. Some observational campaigns confirm this \citep[e.g.,][]{Kim2009ApJ, Husband2013MNRAS}, while others don't \citep{Morselli2014A&A, Mazzucchelli2017ApJ}. The discrepancy in results may arise due to various selection effects, low-number statistics and stochasticity \citep{Buchner2019ApJ}. In addition, the host galaxies of high-redshift quasars tend to have very high star formation rates \citep[SFR; ][]{Trakhtenbrot2017ApJ,Decarli2018ApJ}, indicating high gas content. The reason for this correlation in our model is the requirement of a massive gas reservoir that can feed the SMBH while withstanding AGN feedback; in some cases, this feedback may be positive and induce a part of the observed high SFR. In contrast, other SMBH origin models require the seeds to form in the largest dark matter overdensities. 

\subsection{Other works}

Some recent works have investigated the evolution of SMBH spin and mass from first principles, both at high redshift or during the whole age of the Universe.

\cite{Sesana2014ApJ} used a semi-analytical model of galaxy evolution to track the mass and spin evolution of black holes from high-redshift seeds to the present day. They use several prescriptions linking the large-scale dynamics of the galaxy to that of the accretion flow on to the SMBH. In all of them, the flow inherits some memory of the angular momentum of the whole system, leading to partially aligned accretion. Predictably, then, the calculated SMBH spin is generally high, especially at high redshift, when gas accretion is the dominant mode of mass growth. The authors did not set out to reproduce the observed extreme quasars, but the few examples of individual SMBH mass and spin evolution show very little growth in the first Gyr after the Big bang. At later times, spin tends to decline somewhat, coming into agreement with available low-redshift observational constraints.

\cite{Griffin2019MNRAS} use the {\scshape Galform} semi-analytical galaxy evolution model to track the mass and spin evolution of black holes. They use stellar black hole seeds and include mergers and chaotic cold gas accretion as modes of evolution of SMBH spin. They recover the present-day SMBH mass function and find that SMBHs should generally have rather low spins $a < 0.5$, but they do not reproduce the extreme high-redshift quasars. This is most likely because of limited sample size which only allows them to account for SMBH space densities $> 10^{-7}$~Mpc$^{-3}$ at $z > 6$.

\cite{Zhang2019ApJ} used a model composed of two SMBH growth phases: an initial super-Eddington coherent phase followed by a number of Eddington-limited chaotic episodes. During the latter, the SMBH spin decreases, but the equilibrium values are typically higher than in our models, mainly due to the somewhat less stringent constraint on maximum disc mass, which leads to more common alignment of the disc and the black hole at the early stages of evolution. This led the authors to conclude, in \citep{Zhang2020arXiv}, that a period of super-Eddington growth is necessary to produce the observed high-redshift quasars. In that paper, they added an additional constraint that the currently-observed Eddington ratio of the high-redshift quasars is the maximum allowed in the chaotic accretion periods of SMBH growth, which further constrained the amount of mass obtained during chaotic accretion episodes. Given that individual episodes of SMBH growth are short compared to the total growth time and that the AGN luminosity can vary significantly during each episode, we do not think such a constraint is warranted.

\section{Conclusion} \label{sec:conclusion}

With a simple numerical model, we have shown that extremely massive black holes, such as \Pon\!, can grow from stellar mass seeds via luminous accretion. The only necessary assumptions are that the black hole is fed almost continuously from a turbulent reservoir that can produce a large number of accretion episodes with initial mass transfer rates at least equal to the Eddington rate and initially uncorrelated directions. This chaotic accretion scenario allows the black hole to maintain a low spin and, hence, low radiative efficiency $\epsilon \simeq 0.06$, which promotes much more rapid growth than the canonically-assumed $\epsilon \sim 0.1$. There is not much scope for relaxation of these constraints; in particular, we show that the model would be falsified if a $10^9 \msun$ SMBH were discovered at $z > 9.1$, corresponding to $t = 545$~Myr after the Big bang. By comparing the observed space density of extreme quasars with the simulated space density of Pop-III star formation, we show that only a very small fraction of seed black holes, $\sim 1$ in $4 \times 10^7$, have to encounter favourable growth conditions to produce the observed SMBHs at $z > 6$. The other seeds grow much less efficiently and result in a population that is much more difficult to observe both due to lower mass and lower duty cycle.

\section*{Data Availability}

The code used to produce the figures is available from the corresponding author under reasonable request.

\section*{Acknowledgements}

KZ is funded by the Research Council Lithuania grant no. S-MIP-20-43. Theoretical astrophysics in Leicester is supported by the STFC Consolidated Grant.




\bibliographystyle{mnras}
\bibliography{zubovas} 








\bsp	
\label{lastpage}
\end{document}